\documentclass{article}
\usepackage{spconf,amsmath,graphicx}
\usepackage{multirow}
\usepackage{geometry}

\title{Efficient RE-PARAMETERIZATION RESIDUAL ATTENTION NETWORK FOR NONHOMOGENEOUS IMAGE DEHAZING}
\name{Tian Ye$^{1\dagger}$ \qquad ErKang Chen$^{1\star}$ \qquad XinRui Huang$^{1}$ \qquad Peng Chen$^{1}$    }
\address{$^{1}$School of Ocean Information Engineering,JiMei University \\ }
\begin{document}
%
\maketitle
\begin{abstract}
This paper proposes an end-to-end Efficient Re-parameterization Residual Attention Network(ERRA-Net) to restore the nonhomogeneous hazy image directly.The contribution of this paper mainly has the following three aspects:
1) A novel Multi-branch Attention (MA) block. Spatial
attention mechanism can better reconstruct high-frequency
features, and channel attention mechanism treats the
features of different channels differently.Multi-branch
structure dramatically improves the representation ability
of the model and can be changed into a single path
structure after re-parameterization to speed up the process
of inference. Local Residual Connection allows the low frequency information in the nonhomogeneous area to
pass through the block without processing so that the
block can focus on detailed features.
2) A lightweight network structure. We use cascaded
MA blocks to extract high-frequency features step by step,
and the Multi-layer attention fusion tail combines the
shallow and deep features of the model to get the residual
of the clean image finally.
3)We propose two novel loss functions to help reconstruct
the hazy image Color Attenuation loss and Laplace
Pyramid loss.
ERRA-Net has an impressive speed, processing
1200x1600 HD quality images with an average runtime of
166.11 fps. Extensive evaluations demonstrate that ERSANet performs favorably against the SOTA approaches on
the real-world hazy images.
\end{abstract}
\section{Introduction}
\label{sec:intro}
Outdoor images of Onboard cameras and outdoor
monitoring often deteriorate due to the bad weather, such
as fog and haze; it has dramatically affected the visibility of
images, as shown in fig.1.Restore the image has become a fundamental problem, and people's life
is closely linked. Most early methods are not very good for
non-uniform hazy images. Moreover, they are all based on
the classic atmospheric scattering model, shown as the
following equation.1.
\begin{equation}
    I(x)=J(x)t(x) + A(1-t(x))
\end{equation}
where,$I(x)$ is the hazy image,$J(x)$ is the dehazed
image,$t(x)$ is the medium transmission function and A
is the global atmospheric light. The traditional method
based on the atmospheric scattering model like DCP is
invalid for nonhomogeneous images, as shown in fig.6. The
model that can quickly and accurately remove the haze can
significantly improve the visibility of the image and has a
high application value; however, most of the current
models have heavy parameters and a large number of
calculations, like FFA-Net\cite{qin2020ffa}(Param:4.68M), MSBDN\cite{dong2020multi}
(Param:31.35M), and cannot remove various haze in the
real world quickly. Therefore the obtained dehaze images
are not visually satisfactory, with problems such as color
bias and blurring. We believe that it is valuable to design a
lightweight network model that can quickly repair
nonhomogeneous images. In response to this problem, we
designed the ERRA-Net.Our model consists
mainly of convolutions with a kernel size of 3x3 or 1x1.
This structure gives our model good training potential and
inference performance when the inference time combines
multiple branches into a single branch. Since the model is small and light,
our model can be well deployed on mobile embedded
devices.

\begin{figure*}[htb]

\begin{minipage}[b]{1.0\linewidth}
  \centering
  \centerline{\includegraphics[width=14cm]{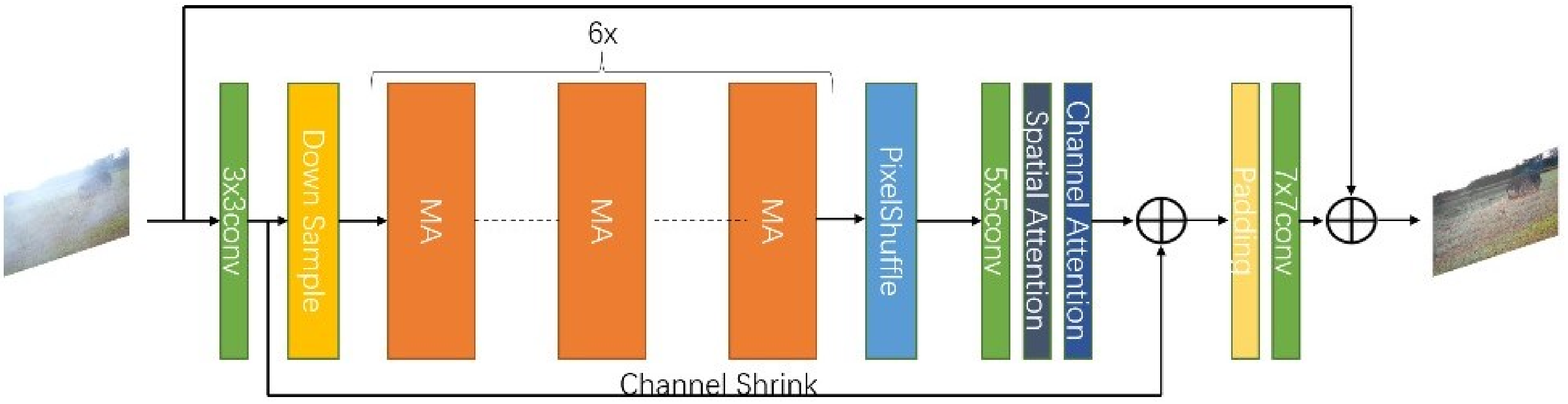}}
\end{minipage}

\caption{The Efficient Re-parameterization Residual Attention Network (ERRA-Net) architecture.}
\label{fig:res}
\end{figure*}

\section{Related Works}
\label{sec:RelatedWorks}
The image dehazing method is to generate a clear
image without haze from the hazy image. There are
two main methods:Prior-based methods and learning-based methods.
\subsection{Prior-based Methods}
 DCP\cite{he2010single} proposed a dark channel prior,but the prior
is unreliable sometime in real-world,like when the color
of objects are too similar to the atmospheric light.
Color Attenuation Prior\cite{he2010single} is a priority based on the
difference between brightness and saturation. Under
the influence of haze, the local image saturation will decrease, the scene’s color will disappear, and the brightness will increase. The saturation and brightness will
form a difference. We can use this difference to estimate the fog concentration:
 \begin{equation}
    A(x)=I^v(x)-I^s(x)
\end{equation}
 $I^v(x)$ and $I^s(x)$ is in the HSV color space.
\subsection{Learning-based Methods}
In recent years, learning-based methods have shined
in image denoising, defogging, and defogging. Large scale dataset (such as RESIDES\cite{li2018benchmarking}) are used for training to
allow the model to learn relevant features to restore
degraded images. Some learning-based methods are
still based on the traditional atmospheric scattering
model.
AOD-Net\cite{li2017aod} estimate the only unknown parameter
in the re-formulated atmospheric scattering model, instead of separately estimating the transmission map
and global atmospheric light,to obtain a better visual
effect.Furthermore, many methods have moved away
from the traditional atmospheric scattering model
and instead use more complex modules and network
structures to obtain better results.FFA-Net[9] proposed a novel Feature Attention module that combines Pixel Attention and Channel Attention and a
Multi-level feature fusion structure based on attention mechanism.FFA-Net\cite{qin2020ffa} showed good results on
the SOTS[8]. MSBDN\cite{dong2020multi} proposed a dense feature fusion module and designed a U-net structure Network based on the principle of boosting and error
feedback.DMPHN\cite{das2020fast} extracts and aggregates features from multiple image blocks in different spaces of the hazy image.


\section{Efficient Re-parameterization Residual Attention Network(ERRA-Net)}
\subsection{Training-time Multi-branch Architecture}
Inspired by RepVGG\cite{ding2021repvgg}, we use a multi-branch
structure during training, as shown in fig2. Multi-branch
Architecture has better training potential than Plain CNNs.
The multi-branch architecture makes the model a lot of An
implicit collection of shallower models\cite{veit2016residual}. Therefore, we
designed the MA Block as the main
backbone of the model.
The Multi-branch of the MA Block
comprises 3x3 convolution, 1x1 convolution, and residual
connection, as shown in fig.3. This parallel multi-branch
design allows the convolution of different receptive fields to
extract features of different scales to better extract features
from non-uniform hazy images. Local Residual
Learning allows high-frequency areas, that is mist or the
information of the clean part passes directly without
processing so that the 1x1 convolution and 3x3 convolution
can better deal with the characteristics of the low-frequency
area (dense fog area) in the picture. We weighed speed and
accuracy ,chose convolution of the kernel size of 3x3
and 1x1. This structure design can improve the
representational ability of the model, and integrate Multi-branch into a single-path through re-parameterization, as  
shown in fig.2, thereby effectively reducing the amount of
calculation. We conducted a detailed ablation experiment,
as shown in the table.2 , to prove the effectiveness of the Block.

\subsection{Attention module}
Our attention module is mainly composed of Spatial
Attention and Channel Attention, focusing on features of
different dimensions and applying attention mechanisms to
different dimensions. 
\subsubsection{Spatial Attention}
In real-world hazy images, object details tend to be high frequency regions, while regions such as object contours
have low frequencies. Therefore, we use 3D MaxPool as the
first step of spatial attention to highlight the detailed
features of objects, and then use 1x1 Conv, ReLU, and 1x1
Conv to adjust to different levels of features, and then
Sigmoid to output 0-1 weight, and spatial attention as
shown in Figure 4. The following equation produces our
proposed spatial attention map:
\begin{equation}
    Y_{sa}=Sigmoid(Conv(Relu(Conv(MaxPool(X_{in})))))
\end{equation}
Wherein,MaxPool is 3d Maxpool.The shape changes from $B\times C\times H \times W$ to $B \times 1 \times H \times W$.
We utilize element-wise mutiplication for input $X_{in}$ and the Spatial Attention map $Y_{pa}$ to get the output of the Spatial Attention(SA) module,finally.

\begin{equation}
 F_{out} = Y_{sa} \otimes x_{in}
\end{equation}

\subsubsection{Channel Attention}
The characteristics of different channels often have
different degrees of importance, so the channel attention
mechanism is necessary. We follow the Channel Attention
design in the FA block\cite{qin2020ffa}.

\subsection{The Efficient Long Skip Connection Attention Structure}
As shown in fig.1, after a 3x3 convolution, the tensor's shape becomes $B \times 64 \times H\times W $, and then a 3x3 convolution with a step size of 2 is used for a downsampling operation ,which will double the receptive field of the subsequent convolution and improve the model's performance. Then, after 6 MA blocks, the features are gradually extracted, and the original resolution is restored after the up-sampling operation. In order to avoid the loss of shallow features, before fusion tails, channel shrink and long-distance residuals are used to introduce shallow features. At this time, the shape is $B \times 16 \times H \times W$, and then it passes through Reflection Padding, and The 7x7 convolution constitutes a fusion tail to aggregate features. The 7x7 convolution has a large receptive field and can make full use of the features. The Reflection Padding operation makes the edges of the feature map better preserved, and the residual of the final result $Y_{res}$ is obtained, and finally, $Y_{res} $ Add input $X_{input}$ to get the final clean image $Y_{clean}$:
\begin{equation}
    Y_{clean} = Y_{res} + X_{input}
\end{equation}

Our multi-layer fusion tail can better combine the shallow and deep features, avoiding the instability caused by the upsampling process and the influence of the large convolution kernel on the image edge restoration effect.

The structure of ERRA-Net is a good balance between
the amount of calculation and accuracy. Under the
premise of ensuring speed, it maximizes every step of
the operation.
We design the network model based on the shallow
structure rather than the deep structure. 

     
     


\subsection{Re-param for Inference-time Model}

\begin{table*}[t]
\centering
\begin{tabular}{|c|l|c|c|c|c|c|c|}
\hline
\multicolumn{2}{|c|}{\multirow{2}{*}{Method}} & \multicolumn{2}{c|}{NH-Haze}     & \multicolumn{2}{c|}{Dense-Haze}  & \multirow{2}{*}{Param} & \multirow{2}{*}{Runtime(fps)@1600x1200} \\ \cline{3-6}
\multicolumn{2}{|c|}{}                        & PSNR            & SSIM           & PSNR           & SSIM            &                        &                                         \\ \hline
\multicolumn{2}{|c|}{(TPAMI)DCP\cite{he2010single}}           & 10.57           & 0.52           & 10.06          & 0.3856          & \textbackslash{}       & \textbackslash{}                        \\ \hline
\multicolumn{2}{|c|}{(TIP)AOD-Net\cite{li2017aod}}         & 14.104          & 0.552          & 13.34          & 0.4244          & \underline{0.002M}        &2598.3                                         \\ \hline
\multicolumn{2}{|c|}{(AAAI)FFA-Net\cite{qin2020ffa}}        & 19.1            & 0.748          & 14.31          & 0.4797          & 4.68M                  & 1.28                                    \\ \hline 
\multicolumn{2}{|c|}{(CVPR)MSBDN\cite{dong2020multi}}          &19.31                 &0.759                & 15.41          & 0.4858          & 31.35M                 &36.84                                         \\ \hline
\multicolumn{2}{|c|}{(CVPRW)DMPHN\cite{das2020fast}}          & 18.184          & 0.745          & 14.01          & 0.4436          & 5.424M                 & 135.46                                  \\ \hline
\multicolumn{2}{|c|}{ERRA-Net(Ours)}           & \underline{19.813} & \underline{0.765} & \underline{15.78} & \underline{0.5154} & 0.3M                   & 166.11                         \\ \hline
\end{tabular}
\caption{Quantitative comparisons with SOTA methods on the real-world dehazing datasets.}
\label{table:1}
\end{table*}


We do not use BN in each branch like RepVGG\nocite{ding2021repvgg},but use the BN layer to stabilize the training effect after the three-way branch,since BN actually destroys the internal features of the sample, we believe that if BN is added to the three branches, the feature connections within the sample will be destroyed more seriously. The ablation experiment also proved our point.We first do the fusion of the three branches,combines 3x3 Conv, 1x1 Conv and identity branch into a 3x3 convolution.
We use $W^k$ to indicate the convolution with size $k$.
The $Y$ denotes the output, and the $X$ denotes the input, and the following formula can describe the output of the multi-branch part:

\begin{equation}
 Y =  BN(W^3\times X + W^1 \times X + X)
\end{equation}
Among them, the identity branch can be expressed as a particular 1x1 convolution, and the 1x1 convolution can be padding with 0 as a 3x3 convolution.\textbf{Please refer to the supplementary materials for the complete derivation of multi-branch fusion.}The whole fusion process is shown in fig.5. About 30 fps can improve the reparameterized ERRA-Net on the original basis (1200x1600 HD quality image on RTX 2080ti).

\begin{figure}[htb]

\begin{minipage}[b]{.48\linewidth}
  \centering
  \centerline{\includegraphics[width=4.0cm]{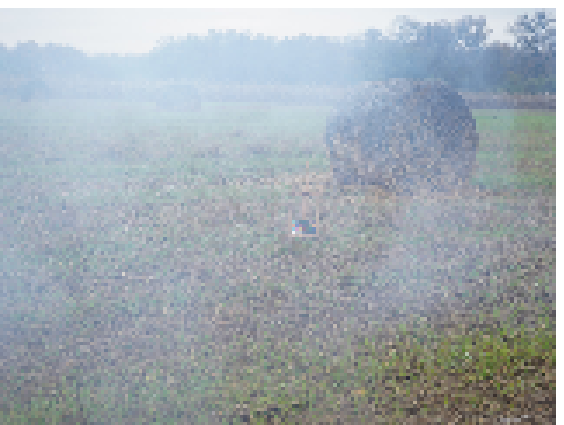}}
  \centerline{(a) Hazy}\medskip
\end{minipage}
\hfill
\begin{minipage}[b]{.48\linewidth}
  \centering
  \centerline{\includegraphics[width=4.0cm]{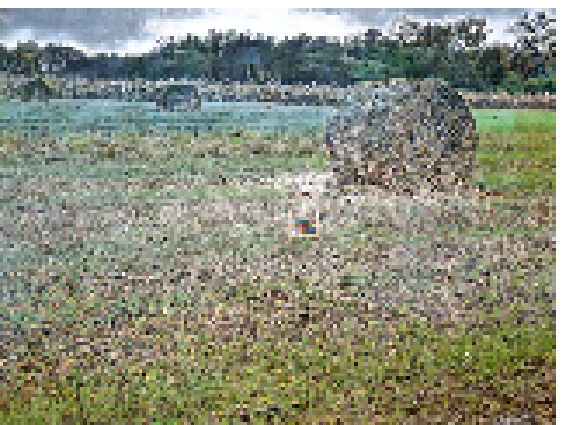}}
  \centerline{(b) DMPHN\cite{das2020fast}}\medskip
\end{minipage}

\begin{minipage}[b]{0.48\linewidth}
  \centering
  \centerline{\includegraphics[width=4.0cm]{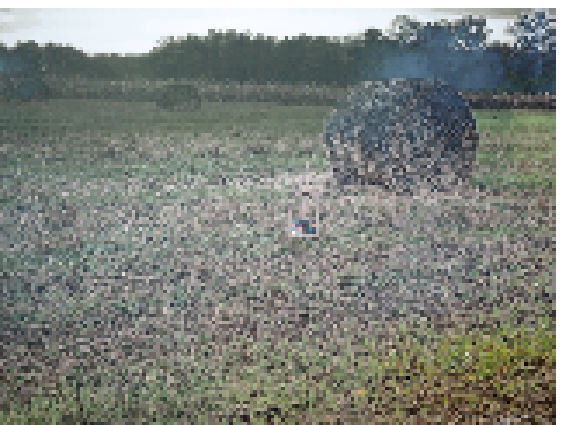}}
  \centerline{(c) FFA-Net\cite{qin2020ffa}}\medskip
\end{minipage}
\hfill
\begin{minipage}[b]{0.48\linewidth}
  \centering
  \centerline{\includegraphics[width=4.0cm]{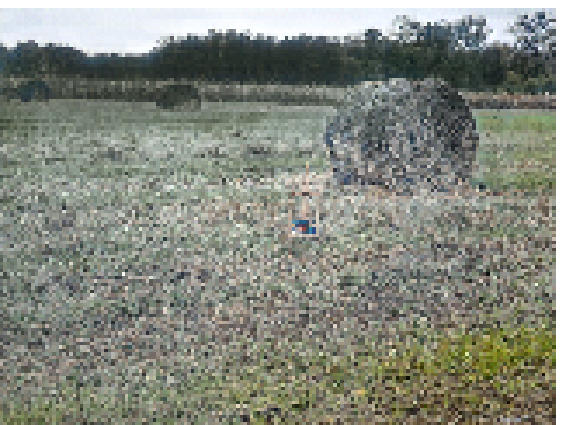}}
  \centerline{(c) MSBDN\cite{dong2020multi}}\medskip
\end{minipage}

\begin{minipage}[b]{0.48\linewidth}
  \centering
  \centerline{\includegraphics[width=4.0cm]{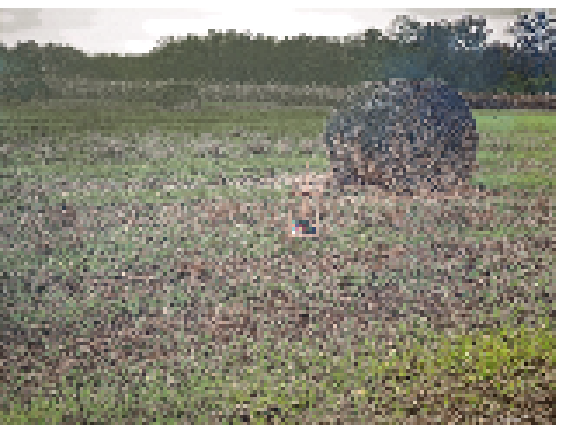}}
  \centerline{(c) Ours}\medskip
\end{minipage}
\hfill
\begin{minipage}[b]{0.48\linewidth}
  \centering
  \centerline{\includegraphics[width=4.0cm]{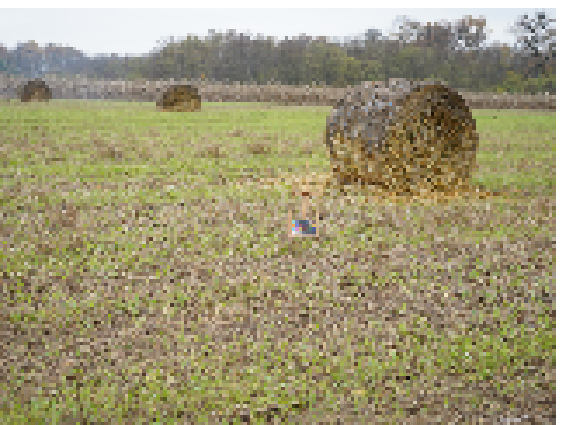}}
  \centerline{(c)Ground Truth}\medskip
\end{minipage}

\caption{Visual comparison on NH-Haze2021\cite{2021ntire} datasets.}
\label{fig:res}
\end{figure}

\section{Loss function}
The loss function we use is:
\begin{equation}
    L_{\Theta} = L1 + \alpha_1 L_{CA} + \alpha_2 L_{laplace}
\end{equation}

Where $\Theta$ denotes the parameters of ERRA-Net,L1 is the basic reconstruction loss ,$L_{CA}$ is Color Attenuation Loss,$L_{laplace}$ is Laplace Pyraimd Loss,and $\alpha_1,\alpha_2$is the balance coefficient, it is usually set to 0.5 and 5. It is worth noting that we did not use the commonly used Perceptual Loss and GAN Loss.

\subsection{Laplace Pyramid Loss}
We designed a novel three-layer Laplace Pyramid to guide the model to better recover high-frequency features as shown in fig.4.The $Y_{output}$ and clean image output by the model will undergo four downsampling operations after Gaussian blur filtering to generate $GD_n,n=1,2,3,4$, Use Bilinear Interpolate to recover the $n$th level image $HR_{n}$ from the $n-1$th level low-resolution image $LR_{n-1}$, and $LP_n = HR_n - LR_{n-1},n=1,2,3$.The obtained $LP_n$ contains the extracted high-frequency features, use the following formula to compute the final $L_{laplace}$ loss:

\begin{equation}
\begin{split}
     L_{laplace}&= \sum_{k=\{1,2,3\}} \frac{1}{N}\sum_{i=1}^{N}\left\|LP_{k}(ERSA(I_{haze}^i))-LP_{k}({I}_{gt}^i)\right\|_{2}^{2} 
\end{split}
\end{equation}

Where $I_{gt}$ stands for ground truth ,and $I_{haze}$ stands for input.

\subsection{Color Attenuation Loss}
 We are inspired by color attenuation prior\cite{colorattenuationprior} and propose Color Attenuation Loss.
The haze concentration $P(x)$ of a haze map can be calculated and expressed by the following formula:
\begin{equation}
 P(x) = \left\|S(I(x)) - V(I(x))\right\|    
\end{equation}
Wherein, $ S (x) $ represents obtains $ I (x) $ saturation, $ V (x) $ represents obtains $ I (x) $ luminance. Under the influence of fog and haze, the saturation of the local area of the picture decreases, while the color of the local scene gradually disappears, while the brightness increases, and the saturation and brightness form a difference. As shown in Figure 7, the difference between the average brightness and the average saturation of the dense fog area is significantly larger than that of the non-fog area. According to this prior, the difference value of brightness and saturation can be used to estimate the concentration of fog, then the supervision of brightness and saturation respectively can get the loss function describing the concentration of haze——$L_{CA}$:
\begin{equation}
\begin{split}
L_{CA}=\alpha\frac{1}{N} \sum_{i=1}^{N}\left\| S(I_{g t}^{i})- S(ERSA\left(I_{\text {haze }}^{i})\right)\right\|
\\
+\beta \frac{1}{N} \sum_{i=1}^{N}\left\| V(I_{g t}^{i})- V(ERSA\left(I_{\text {haze }}^{i})\right)\right\|_2^2
\end{split}
\end{equation} 
Where $\alpha$ and $\beta$ is the coefficient to regulate the importance of the difference of brightness and saturation,$I_{gt}$ stands for ground truth ,and $I_{haze}$ stands for input. The values of $\alpha$ and $\beta$ are respectively 1 and 0.5 in our experiment.


\section{Experiment}
\subsection{Datasets and Metrics}


We chose real-world datasets to train and evaluate our ERRA-Net.We adop the O-Haze,the I-Haze\cite{I-HAZE_2018}  and NH-Haze\cite{ancuti2020ntire}  as the training set.Our training set has a total of 130 pictures.The Dense-Haze\cite{Dense-Haze_2019} and NH-Haze\cite{2021ntire} 2021  as two benchmark for single image dehazing.To evaluate the performance of our ERRA-Net,we adopt the Peak Signal to Noise Ratio(PSNR) and the Structural Similarity index (SSIM) as the evaluation Metric.The SSIM and PSNR quantify the structural similarity and reconstruction quality of the output image with the respective ground truth image.

\subsection{Training Settings}
We train the ERRA-Net in RGB channels, augment the training dataset with randomly rotated and horizontal flip.The 2 image patches with the size 128x128 are extracted as $X_{input}$ of ERRA-Net.The network is trained for $6\times10^3$ steps on the training dataset.We use Adam optimizer,where $\beta1$ and  $\beta2$ are 0.9 and 0.999,respectively.
The initial learning rate is set to $6\times10^-4$,and we adopt the Cyclical Learning Rate strategy\cite{smith2017cyclical}to adjust the learning rate from the initial to $1.2\times10^-3$, where the step size is 10. Pytorch\cite{automatic} was used to implement our models with 2 RTX 2080Ti GPU.


\subsection{Comparison with State-of-the-art Methods}
We compare our ERRA-Net with SOTA methods on Dense-Haze and NH-Haze2021 datasets.As shown in Table.1,we can observe:
(1)Our ERRA-Net outperforms all SOTA methods with 19.813dB PSNR and 0.765 SSIM on NH-Haze2021\cite{2021ntire}.
(2)Our ERRA-Net also outperforms all SOTA methods with 15.731dB PSNR and 0.5123 SSIM on Dense-Haze dataset.
(3)We compare our ERRA-Net with SOTA methods on the quality of restored images on NH-Haze2021\cite{2021ntire} and Dense-Haze\cite{Dense-Haze_2019} ,which are presented in Fig.8-11.Obviously,our method generates the most clean and natural images,compared to other methods.(4)Apart from proper dehazing results,it's to be noted that our ERRA-Net is lightweight and efficient models, checkpoints of our model take 1.20 MB on disk, and the total parameters of our model is only 302.513K.
\subsection{Ablation Study}
In this section,we verify the effectiveness of different modules in ERRA-Net,the Color Attenuation loss and the Laplace Pyramid loss.

We first construct our base model of ERRA-Net as the baseline.The MA block of base Model only consists of Multi-branch and activation function.We add the different component into base model as:(1)\textbf{base + BN + AM}:Add the BN layer,Spatial Attention Module and Channel Attention Module into baseline.(2)\textbf{base + BN + AM + LR}:Add the BN layer,Attention Module and Local Residual into baseline.Experiments show that the addition of the BN layer and attention mechanism after multiple branches can greatly improve the performance of the model,as shown in Table 2.

We also verified the effectiveness of the two newly proposed loss functions,as shown in Table 2.We found that the $L_{laplace}$  loss is very good for the improvement of PSNR and SSIM.
Table 3 shows the effect that we set for the BN layer in the MA block on the model performance. Too many BN layers in a block will destroy the internal feature relationship of the sample and reduce the performance. We are adding the BN layer after Multi-branch will stabilize the training of the model.

\begin{table}[]
\centering
\begin{tabular}{|c|cc|}
\hline
Model/Loss                   & PSNR   & SSIM  \\ \hline
base                    & 18.797 & 0.733 \\
base+BN+AM              & 19.532 & 0.754 \\
base+BN+AM+LR           & 19.813 & 0.765 \\ \hline
$L1$                      & 19.21  & 0.746 \\
$L1 + L_{ca}$              & 19.34  & 0.741 \\
$L1 + L_{ca} + L_{laplace}$ & 19.813 & 0.765 \\ \hline
\end{tabular}
\caption{Ablation Study on ERRA-Net.}
\label{table:2}
\end{table}

\begin{table}[]
\centering
\begin{tabular}{|c|c|c|}
\hline
Each branch with BN & 18.627 & 0.688 \\ \hline
Multi-branch with BN  & 19.813 & 0.765 \\ \hline
\end{tabular}
\caption{Adding a BN layer to each branch will reduce the model's performance.}
\label{table:3}
\end{table}

\section{Conclusion}
This paper propose an efficient Re-parameterization Residual Attention Network and demonstrated its strong power in nonhomogeneous image restoration.Although our model has a simple structure, it performs well on real-world haze images, with very few parameters and breakneck speed. We hope to apply it to other low-level vision tasks in the future.
\section{acknowledgments}
This research was funded in part by the Education Department of Fujian Province (grant numbers JAT190301), in part by the Foundation of Jimei University (grant numbers ZP2020034).

\begin{figure*}[]

\begin{minipage}[b]{.3\linewidth}
  \centering
  \centerline{\includegraphics[width=4.0cm]{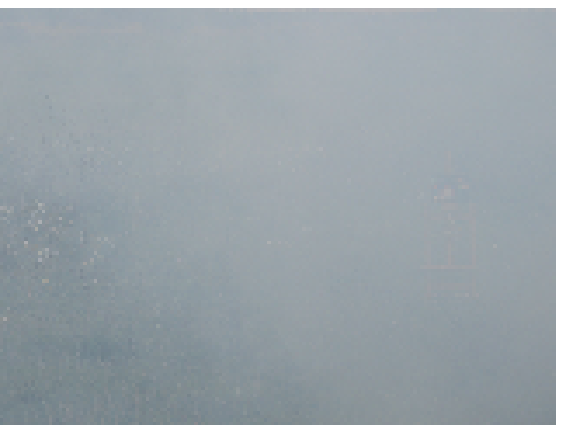}}
  \centerline{(a) Hazy}\medskip
\end{minipage}
\begin{minipage}[b]{.3\linewidth}
  \centering
  \centerline{\includegraphics[width=4.0cm]{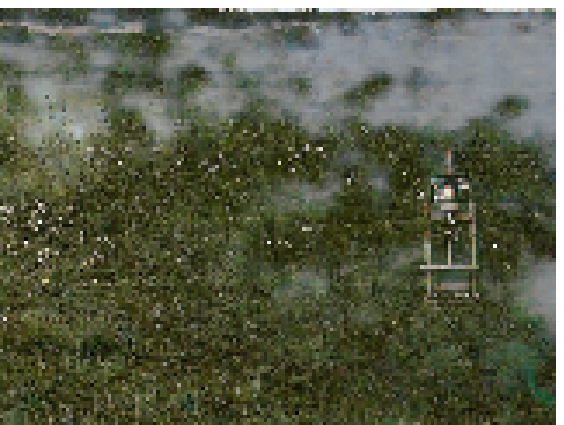}}
  \centerline{(b) DMPHN\cite{das2020fast}}\medskip
\end{minipage}
\begin{minipage}[b]{0.3\linewidth}
  \centering
  \centerline{\includegraphics[width=4.0cm]{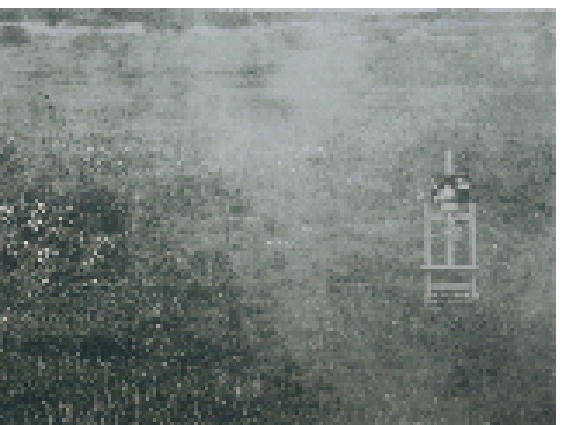}}
  \centerline{(c) FFA-Net\cite{qin2020ffa}}\medskip
\end{minipage}

\begin{minipage}[b]{0.3\linewidth}
  \centering
  \centerline{\includegraphics[width=4.0cm]{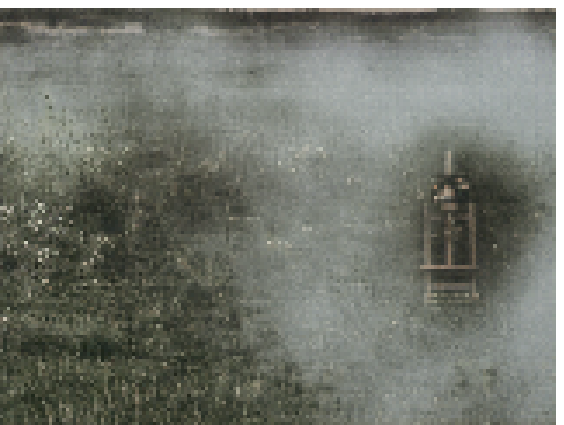}}
  \centerline{(c) MSBDN\cite{dong2020multi}}\medskip
\end{minipage}
\begin{minipage}[b]{0.3\linewidth}
  \centering
  \centerline{\includegraphics[width=4.0cm]{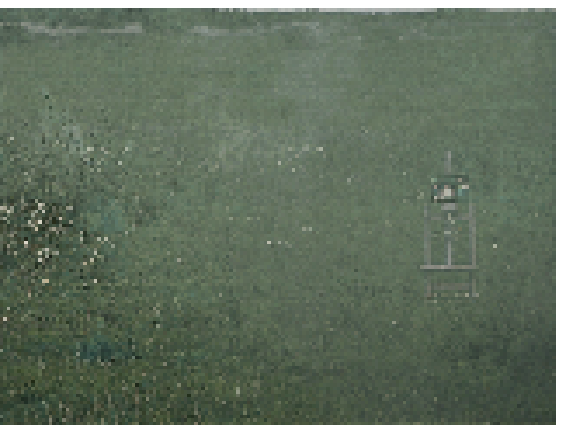}}
  \centerline{(c) Ours}\medskip
\end{minipage}
\begin{minipage}[b]{0.3\linewidth}
  \centering
  \centerline{\includegraphics[width=4.0cm]{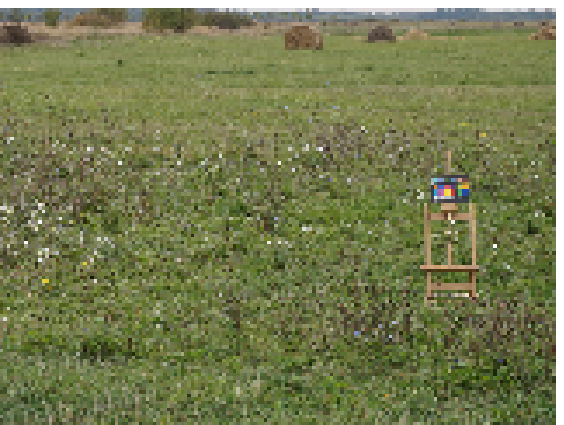}}
  \centerline{(c)Ground Truth}\medskip
\end{minipage}

\caption{Visual comparison on NH-Haze2021\cite{2021ntire} datasets.}
\end{figure*}

\begin{figure*}[]

\begin{minipage}[b]{.3\linewidth}
  \centering
  \centerline{\includegraphics[width=4.0cm]{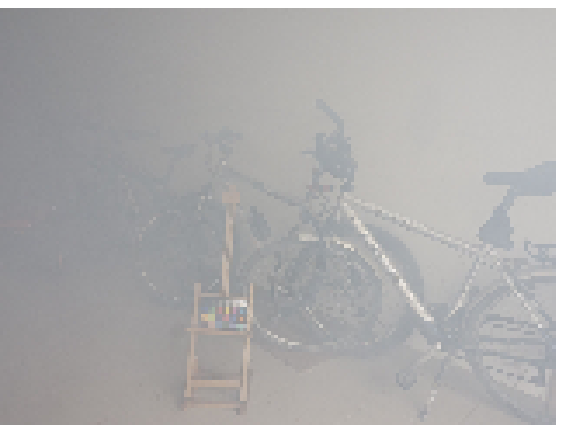}}
  \centerline{(a) Hazy}\medskip
\end{minipage}
\begin{minipage}[b]{.3\linewidth}
  \centering
  \centerline{\includegraphics[width=4.0cm]{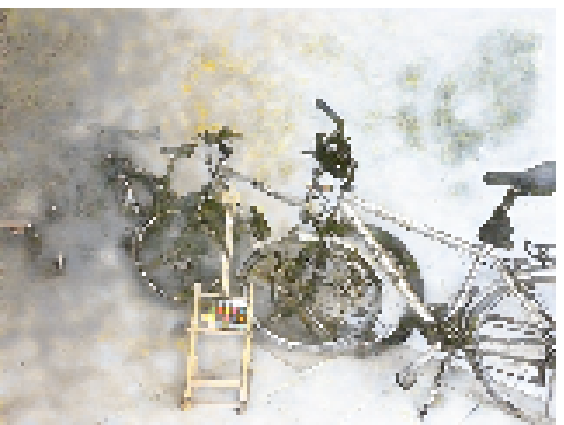}}
  \centerline{(b) DMPHN\cite{das2020fast}}\medskip
\end{minipage}
\begin{minipage}[b]{0.3\linewidth}
  \centering
  \centerline{\includegraphics[width=4.0cm]{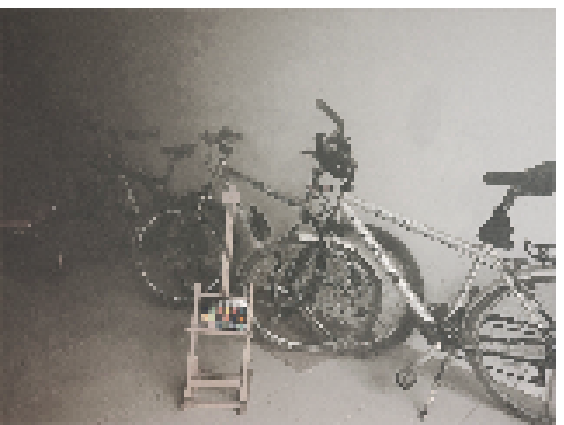}}
  \centerline{(c) FFA-Net\cite{qin2020ffa}}\medskip
\end{minipage}

\begin{minipage}[b]{0.3\linewidth}
  \centering
  \centerline{\includegraphics[width=4.0cm]{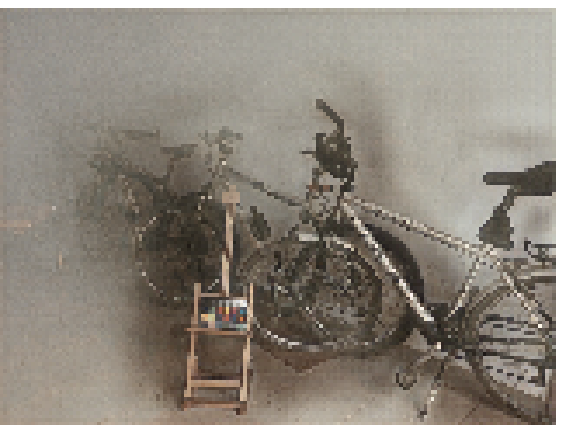}}
  \centerline{(c) MSBDN\cite{dong2020multi}}\medskip
\end{minipage}
\begin{minipage}[b]{0.3\linewidth}
  \centering
  \centerline{\includegraphics[width=4.0cm]{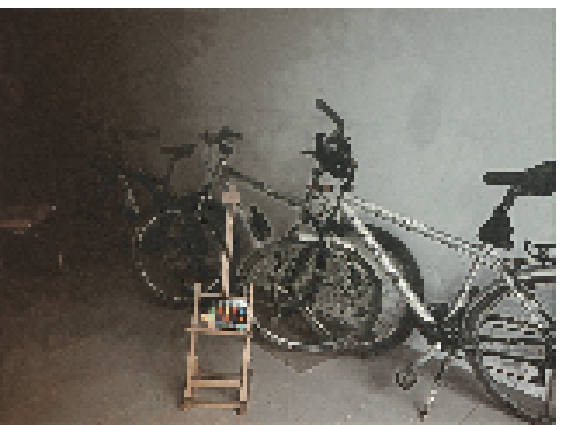}}
  \centerline{(c) Ours}\medskip
\end{minipage}
\begin{minipage}[b]{0.3\linewidth}
  \centering
  \centerline{\includegraphics[width=4.0cm]{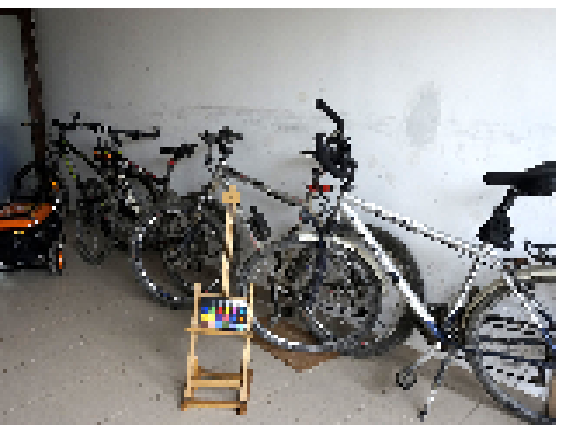}}
  \centerline{(c)Ground Truth}\medskip
\end{minipage}

\caption{Visual comparison on NH-Haze2021\cite{2021ntire} datasets.}
\end{figure*}

\begin{figure*}[]

\begin{minipage}[b]{.3\linewidth}
  \centering
  \centerline{\includegraphics[width=4.0cm]{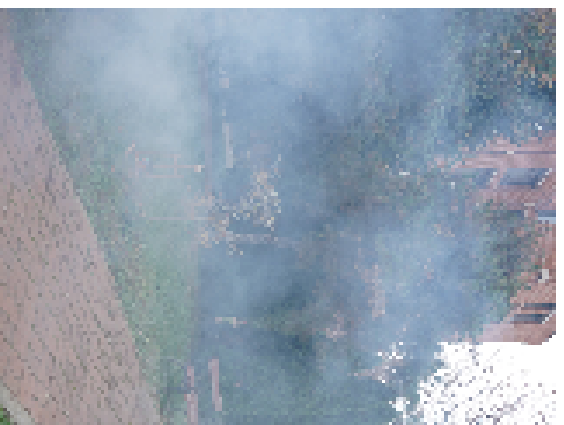}}
  \centerline{(a) Hazy}\medskip
\end{minipage}
\begin{minipage}[b]{.3\linewidth}
  \centering
  \centerline{\includegraphics[width=4.0cm]{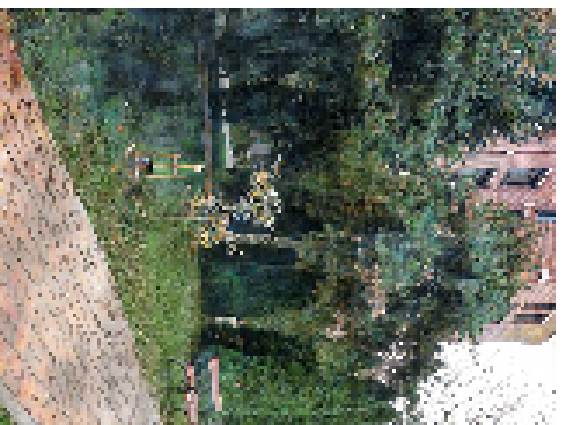}}
  \centerline{(b) DMPHN\cite{das2020fast}}\medskip
\end{minipage}
\begin{minipage}[b]{0.3\linewidth}
  \centering
  \centerline{\includegraphics[width=4.0cm]{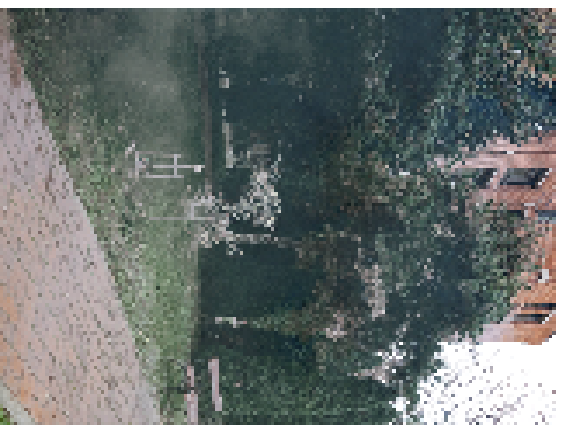}}
  \centerline{(c) FFA-Net\cite{qin2020ffa}}\medskip
\end{minipage}

\begin{minipage}[b]{0.3\linewidth}
  \centering
  \centerline{\includegraphics[width=4.0cm]{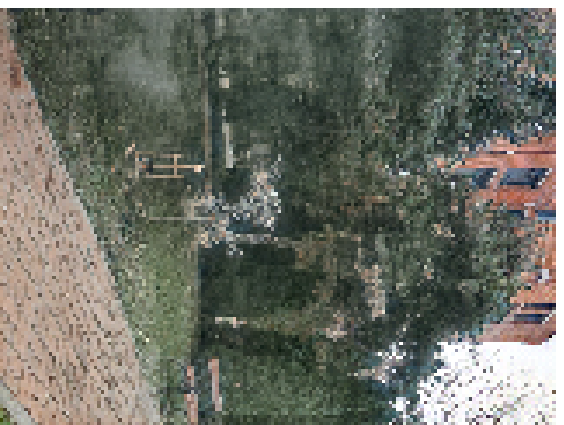}}
  \centerline{(c) MSBDN\cite{dong2020multi}}\medskip
\end{minipage}
\begin{minipage}[b]{0.3\linewidth}
  \centering
  \centerline{\includegraphics[width=4.0cm]{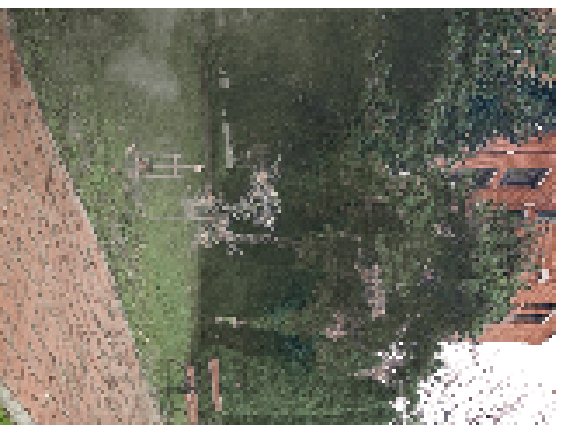}}
  \centerline{(c) Ours}\medskip
\end{minipage}
\begin{minipage}[b]{0.3\linewidth}
  \centering
  \centerline{\includegraphics[width=4.0cm]{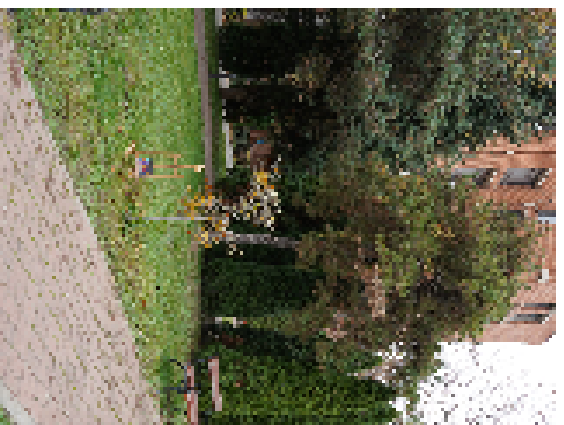}}
  \centerline{(c)Ground Truth}\medskip
\end{minipage}

\caption{Visual comparison on  NH-Haze2021\cite{2021ntire} datasets.}
\end{figure*}

\bibliographystyle{IEEEbib}
\bibliography{Template}

\end{document}